\documentclass[]{article}
\usepackage{graphicx}
\usepackage{amssymb}
\usepackage{amsmath}
\usepackage{txfonts}
\usepackage{natbib}
\usepackage{longtable}
\bibpunct{(}{)}{;}{a}{}{,}

\begin{document}
\title{\hbox{Intermediate mass black holes - observational challenges} \vspace{0.1in} \large{Lecture notes for the Croatian Black Hole School\\Trpanj, Pelje\v{s}ac, Croatia \\ June 21-25, 2010} \vspace{1in}}
\author{Mario Pasquato\\
Dipartimento di Fisica E. Fermi,\\
Universit\`a di Pisa,\\
Largo Bruno Pontecorvo 3,\\
I-56127 Pisa, Italy}
\date{\phantom{:-)}}
\maketitle

\thispagestyle{empty}
\newpage
\thispagestyle{empty}
{\noindent \textbf{Abstract:} Black holes are fascinating objects. As a class of solutions to the Einstein equations they have been studied a great deal, yielding a wealth of theoretical results. But do they really exist? What do astronomers really mean when they claim to have observational evidence of their existence? To answer these questions, I will focus on a particular range of black-hole masses, approximately from ${10}^2$ $M_\odot$ to ${10}^4$ $M_\odot$. Black holes of this size are named Intermediate Mass Black Holes (IMBHs) and their existence is still heavily disputed, so they will be perfect for illustrating the observational challenges faced by a black hole hunter.\vspace{1in}\\}

{\noindent \textbf{Sa\v{z}etak:} Crne rupe su fascinantan objekt. One su, kao klasa rje\v{s}enja Einsteinovih jednad\v{z}bi, bile predmetom mnogih istra\v{z}ivanja koja su dovela do obilja teorijskih rezultata. No postoje li one zaista? \v{S}to astronomi to\v{c}no misle kada turde da imaju opservacijske dokaze da one postoje? Da bih odgovorio na ova pitanja usredoto\v{c}it \'cu se na odredeni raspon mase crnih rupa, onaj izmedu ${10}^2$ $M_\odot$ i ${10}^4$ $M_\odot$. Crne rupe ovih dimenzija nazivaju se crne rupe srednjih masa (IMBHs) i njihovo je postojanje jo\v{s} uvijek na\v{s}iroko osporavano, tako da \'ce one biti savr\v{s}ene za ilustriranje opa\v{z}a\v{c}kih izazova s kojima se suo\v{c}ava lovac na crne rupe.\vspace{1in}\\}

{\noindent \textbf{Sommario:} I buchi neri sono oggetti affascinanti. Come classe di soluzioni delle equazioni di Einstein, sono stati studiati approfonditamente, producendo una messe di risultati teorici. Ma esistono davvero? Che cosa intende dire un astronomo quando afferma di avere delle prove osservative della loro esistenza? Per rispondere a queste domande, mi concentrer\`o sull'intervallo di masse approssimativamente compreso tra ${10}^2$ $M_\odot$ e ${10}^4$ $M_\odot$. I buchi neri di queste dimensioni vengono chiamati buchi neri di massa intermedia (IMBHs) e a tutt'oggi si discute ancora se esistano o meno. Saranno quindi perfetti per illustrare le difficolt\`a osservative incontrate da un cacciatore di buchi neri.}
\newpage
\section{Introduction}
\label{Sect:Introduction}
Intermediate Mass Black Holes (IMBHs) have masses in the range ${10}^2$ -- ${10}^4$ $M_{\odot}$.
Stellar-mass black holes are the product of normal stellar evolution.  Tipically they weigh several solar masses but some have been found heavier than $10$ $M_{\odot}$ \citep[see][]{Cr2008}. To the best of my knowledge the record to date is $15$ $M_{\odot}$ \citep[][]{Bu2007}.
The so-called supermassive black holes, instead, weigh billions of solar masses and have been found in the centers of galaxy bulges, and they are thought to power much of the high-energy activity going on there. \citep[see][for reviews]{Ko1995, Fe2005}.
IMBHs live in-between these two extremes and are actually a promising bridge between them, as they are speculated to constitute the building blocks of supermassive black holes.

\noindent Before plunging into the mysteries of IMBHs we shall deal for a while with stellar mass and supermassive BHs.

\subsection{Stellar-mass black holes}
The formation mechanism of stellar-mass black holes is reasonably understood, as it is based on the time-tested theory of stellar evolution. Black holes are basically what is left after stellar death. How a star dies is dependent on its mass at the time the nuclear reactions capable of sustaining it against the pull of gravity stop. During the giant-branch phase of its evolution, a star typically looses mass in variable amounts, so its final mass is not precisely constrained. Anyway, when nuclear reactions stop, the matter constituting the star collapses, unless other supporting mechanisms intervene. Electron degeneracy pressure is one of these mechanisms, and if the mass of the collapsing star is small enough it will win over gravity and the star will become a white dwarf.
If electron degeneracy pressure yields to gravity, the matter is converted into neutrons and the collapse continues. Neutron degeneracy pressure is the next barrier: if the mass is not high enough, the collapse is stopped by it and a neutron star forms. If the mass is high enough, no known mechanism can stop the collapse and the star is expected to shrink until it is contained within its own Schwarzschild radius, thus becoming a black-hole. 
Most evidence about the existence of stellar-mass black holes comes from X-ray binaries. In these objects, an accretion disk forms and becomes hot enough to emit in the X-ray band \citep[see][]{Mc2003}. X-ray emission plays a crucial role in the context of IMBHs as well, and will be covered in more detail later.

\subsection{Supermassive black holes}
Supermassive black holes were first postulated as engines capable of sustaining the high-power emission of quasars \citep[e.g.][]{Fe2005}. In the local universe they have been observed in the bulges of several galaxies either thanks to the accreting hot gas or indirectly by their effects on stellar kinematics. Also in the bulge of our own Galaxy, the Milky Way, a supermassive black hole was found. It is the supermassive black-hole with the best measurement of mass to date, as the orbits of resolved stars were used to estimate the black hole mass with $\approx 10\%$ uncertainty \citep[][]{Gh2003}.

\subsection{Why focus on IMBHs?}
We have already stated that IMBHs naturally bridge a gap between two observational regimes where relatively firm detections exist.
There are further reasons that make this class of black holes interesting.\\ First of all, IMBHs are likely good candidates as gravitational radiation sources. If they are to be found in a large fraction of massive star clusters (see below why it is thought that star clusters might contain IMBHs), they are likely hundreds of objects in the Local Group alone. In the dense environment of a star cluster, close encounters with stellar-mass black holes and other massive objects would be quite frequent, leading to a good potential for gravitational wave emission. Estimates for LISA \citep[][]{Fr2006} or a similar space-borne interferometric detector of gravitational radiation predict tens of IMBH in-spiral events per year. Somewhat lower numbers are expected for ground based interferometers such as VIRGO or LIGO \citep[see][]{Mi2002b}, which work in a different frequency range and would preferentially reveal black-hole merging events.\\ In a typical galaxy, thousands of star clusters which could potentially contain an IMBH are present. Dynamical friction due to gravitational encounters with stars could lead to an infall of star clusters (with the associated IMBHs) to the galactic bulge. Actually, so-called nuclear star clusters have recently been observed in the bulges of several galaxies. If these where to contain IMBHs they could provide a way to seed super-massive black holes, suggesting an explanation for the formation of the latter \citep[][]{Eb2001, Vo2003, Ta2009}.\\ More generally, IMBHs would have a strong effect on star-cluster dynamics, as they can play a crucial role in the gravothermal collapse of its core (core collapse), basically halting it. Core collapse is one of the processes resulting from the collisional evolution of clusters of stars. Stars in a star cluster evaporate as they diffuse in phase space, attaining velocities that exceed the cluster escape velocity. This way, energy is removed from the cluster, which contracts its core and moves its most massive stars to the center, in order to balance this effect (mass segregation). The three processes of evaporation, core collapse and mass segregation are all linked together in the dynamical evolution of a cluster. The amount of mass segregation attained by stars in a relaxed star cluster is also heavily dependent on the presence of an IMBH \citep[][]{Gi2008}, which, incidentally provides a detection method \citep[][]{Pa2009}.\\ The density and velocity dispersion cusps \citep[][]{Ba1976, No2006, No2007} that have been observed in many star clusters could also be explained by the presence of an IMBH.\\ Finally, IMBHs would also naturally explain the so-called Ultra-Luminous X-ray sources (ULXs), X-ray emitting objects which appear to radiate above the Eddington limit for a stellar-mass black hole \citep[e.g.][]{Ma2000}.

\section{IMBH formation}
\label{Sect:IMBHformation}
Several scenarios for forming black holes in the ${10}^2$ -- $10^4$ $M_{\odot}$ mass range have been proposed.
\begin{itemize}
\item Population III stars, i.e. the first stellar structures to form from primordial-composition gas after the Big Bang, are predicted to be very massive. Their fast evolution can end with a black-hole remnant of several hundreds of solar masses \citep[][]{Ma2001, Sc2002},
\item Simulations of dense clusters with young stars show that they may merge in a runaway cascade, producing a very massive star \citep[][]{Po2004}. Such a star may produce a large black hole when it dies,
\item Lower-mass black holes may merge and produce IMBHs in sufficiently high density environments. Dense star clusters \citep[][]{Mi2002} may thus produce IMBHs, if the accreting object is not kicked out by gravitational wave recoil,
\item Primordial black holes in the intermediate-mass range may form in the early evolution of the universe, according to certain variants of inflationary cosmology \citep[][]{Ka2008}
\end{itemize}

Two out of four mechanisms involve dense environments, so the search for IMBHs is tied to the physics of star clusters in a natural way. Two or more formation routes can be active together, leading to a composite IMBH population.

\subsection{Formation from population III stars}
\subsubsection{Cooling and fragmentation of primordial gas}
Population III stars are extremely metal poor, because they are formed from primordial composition gas. It is the physics of this gas, in particular the cooling mechanisms, which determines how discrete clumps of gas form and collapse to become population III protostars. How the gas cools sets the final mass of the clumps produced, eventually setting the mass of population III stars.
Studies on this topic \citep[][]{Ho1953, Me1965, Me1965b, Lo1976} show that cooling is inefficient in primordial-composition gas, which implies that population III stars are characterized by much higher masses than those of stars forming in the local universe.
With stellar masses as high as hundreds of solar masses, IMBHs can easily be produced as remnants.

A gas chunk of mass $M$ undergoes gravitational collapse if the typical thermal energy $kT$ of a particle is smaller than its binding energy:
\begin{equation}
 \label{JeansCondition}
kT \lesssim \frac{GMm}{R}
\end{equation}
where $R$ is the size of the chunk and $m$ the particle's mass. Introducing the density $\rho \approx M/R^3$ we obtain
\begin{equation}
\label{JeansMass}
M > M_J \approx (kT/Gm)^{3/2} \rho^{-1/2} 
\end{equation}
where $M_J$ is the Jeans mass of the gas for a given density $\rho$ and temperature $T$.
A gas cloud of mass $M \gg M_J$ will split into smaller fragments of mass $\approx M_J$, which collapse separately.
As each chunk contracts, its density and temperature change, and so does $M_J(T, \rho)$. Increases in $M_J$ lead to no further fragmentation, while a decreasing $M_J$ makes the collapse of smaller chunks possible.

If the chunk of gas is unable to efficiently cool, the collapse leads to a temperature increase. The extreme case is adiabatic collapse, where
\begin{equation}
\label{AdiabaticPressureDensity}
T \propto \rho^{\gamma - 1}
\end{equation}
where $\gamma = c_p/c_v = 5/3$ in the case of a purely monoatomic gas.
By substituting Eq.~\ref{AdiabaticPressureDensity} into Eq.~\ref{JeansMass}, we obtain
\begin{equation}
\label{MJwithrho}
M_J \propto \rho^{{(3 \gamma - 4)}/{2}} \propto \rho^{1/2}
\end{equation}
where the last relation holds for $\gamma = 5/3$. 
Density increases during collapse and $M_J$ increases with its square root, therefore no further fragmentation takes place.

The opposite case is that of isothermal collapse, where any heat produced by collapse is immediately dissipated.
Equation \ref{JeansMass} then yields
\begin{equation}
\label{MJwithrhoisothermal}
M_J \propto \rho^{-1/2}
\end{equation}
so the Jeans mass is decreasing as collapse proceeds, and further fragmentation is always possible.
The more efficient the cooling mechanisms, the smaller the mass of the clumps of gas produced at the end of fragmentation.

\subsubsection{Gas cooling mechanisms}
Understanding the physics of gas cooling is crucial to determine the typical mass of the protostars that will form. A collapsing cloud of gas relies mainly on radiative cooling to eliminate the heat produced by the collapse. Radiative cooling depends on the radiative de-excitation of atoms and molecules excited by collisions.

Solar-composition gas contains metals, which are extremely efficient coolants.
To name a few atoms\footnote{In the following we make use of the spectroscopic notation, where $A$~$I$ indicates the neutral atom of element $A$, $A$~$II$ the singly ionized atom, $A$~$III$ the doubly ionized, and so on.}, $C$~$II$, $Ne$~$II$, $Si$~$II$, $N$~$II$, $O$~$II$, $O$~$III$, have transitions which span the energy range $2.5 \times {10}^{-3}$ $eV$ - $5.35$ $eV$. This corresponds temperatures from $\approx 30$ $K$ to $\approx 6.2 \times {10}^{4}$ $K$, which insures the gas is able to efficiently cool over a wide range of temperatures. This entails a decreasing Jeans mass until about $1 M_\odot$ clumps form \citep[e.g.][]{Br2004}. Primordial gas contains virtually no metals, and thus is must cool through less efficient channels. Also dust, which plays a role in the cooling of solar-composition gas, cannot be formed in the absence of metals. The current understanding is that molecular hydrogen, $H_2$, is the main coolant in primordial gas \citep[e.g.][]{Br2002}. This comparatively inefficient cooling leads to protostar masses of $\approx 100$ $M_\odot$ \citep{Br2002, Br2004, Gl2005, No2008}.

\subsubsection{Evolution of primordial-composition stars}
As shown, population III stars are predicted to be very massive based on the physics of primordial gas cooling.
Do they loose lots of mass during their evolution, or can we trust them to keep it until death, allowing the birth of an IMBH?
The massive $O$- and $B$-type stars in the local universe lose mass through radiatively driven winds \citep[e.g.][]{Lu1970, Ca1975, Ca1979, Ab1980}. The force per unit mass $f$ exerted on the gas in the outskirts of such a star depends on the opacity $\kappa$ of the gas and on the radiation flux $\phi$ coming from the core of the star as
\begin{equation}
\label{RadPressOpa}
f = \frac{\kappa \Phi}{c}
\end{equation}
The ratio $f/g$ of radiation pressure acceleration to surface gravitational acceleration determines how much mass loss takes place.
The opacity $\kappa$ is an increasing function of metal content. In a population III star, opacity and thus mass-loss are expected to be much lower than in a solar-metallicity star.

\cite{Vi2001} compute the mass-loss rate of stars for various metallicities and stellar masses. They find that a $60 M_{\odot}$ star with $Z = {10}^{-2} Z_{\odot}$ loses about ${10}^{-7} M_{\odot}/{yr}$ via radiatively driven winds.
A Population III star living for $\approx {10}^7$ $yr$ contains even less metals, leading to a cumulative mass-loss of less than $1 M_{\odot}$ over its lifetime. The mass-loss is thus negligible and the mass stored in the star is fully available to produce an IMBH.
 
Simulations show that non-rotating population III stars with masses in the range from $50$ to $140$ $M_{\odot}$ collapse directly to a black hole, while between $140$ and $260$ $M_{\odot}$, explosion as a Pair-Instability SuperNova (PISN) is expected \citep[][]{He2002, Gl2008}. Masses higher than $260$ $M_{\odot}$ again form a black-hole. Formation of IMBHs from population III stars predicts the existence of a \emph{zone of avoidance} in mass corresponding to the mass range where a PISN occurs, as this kind of supernova is expected to leave no remnant.

\subsection{IMBH formation from VMSs obtained by stellar merging}
The runaway merging of high-mass main-sequence stars in a dense young cluster is a candidate mechanism for IMBH formation. Early on, \cite{Co1967} studied the issue analytically, and in the last decades, extensive numerical simulations of young clusters have become widespread \citep[][]{Fr2006}. These studies show that a Very Massive Star (VMS) forms if the lifetime of massive main-sequence stars ($\approx 3$ $Myr$) is longer than the typical time it takes for the runaway stellar merging to take place. Collisional stellar coalescence has to be fast enough to occur before the explosion of the first supernovae, which cause a sudden expansion of the cluster, reducing the number of physical collisions between stars. If a VMS is produced, its evolution likely leads to an IMBH remnant. Some high-mass stars that have been observed in the local universe, such as the Pistol star \citep[e.g., see][]{Fi1998} could be VMSs produced from such a stellar merging process \citep[][]{Fr2006a}. In the following we will look more closely to the simple argument of \cite{Co1967}.

\subsubsection{A simple model of runaway merging}
\label{CollMe}
Two stars of radii $R_1$, $R_2$ and of masses $M_1$, $M_2$, with typical relative velocity at infinity $V_\infty$, have a cross section for physical collisions in the form:
\begin{equation}
 \label{crosssection}
\sigma = \pi {(R_1 + R_2)}^2 \left(1 + \frac{2 G (M_1 + M_2)}{(R_1 + R_2) V^2_\infty} \right)
\end{equation}
which differs from the purely geometrical cross-section because of the velocity-dependent expression enclosed in the rightmost bracket, which takes into account gravitational focusing. Basically, stars bend each other's orbit via gravitational attraction, resulting in an enhanced collision cross-section. The ratio between the relative velocity at infinity and the escape velocity at contact
\begin{equation}
 \label{escape}
V_E = \sqrt{ \frac{2 G (M_1 + M_2)}{(R_1 + R_2)} }
\end{equation}
determines whether gravitational focusing is important.
Eq.~\ref{crosssection} can be rewritten as
\begin{equation}
 \label{crosssectionVe}
\sigma = \pi {(R_1 + R_2)}^2 \left(1 + \frac{V^2_E}{V^2_\infty} \right)
\end{equation}
so the two limiting cases are $V_E \ll V_\infty$, where gravitational focusing can be neglected, and $V_E \gg V_\infty$ where gravitational focusing is dominant.

For simplicity we will neglect any mass lost in stellar collisions, so all mass ends up in the star resulting from the collision. Let $n$ be the number density of stars in the cluster and let us use the velocity dispersion $\langle v \rangle$ of cluster stars to estimate $V_\infty$. A given star collides with
\begin{equation}
 \label{Nencounters}
\frac{dN}{dt} = n \sigma \langle v \rangle
\end{equation}
other stars per unit time. If $m$ is their average mass, the star's mass evolves as:
\begin{equation}
 \label{MassEvolution}
\frac{dM}{dt} = m n \sigma \langle v \rangle
\end{equation}
where $M$ is the mass of the accreting star.

The radius of a star scales approximately as $R \propto M^\beta$. Escape velocity at the star's surface scales as $V^2_E = M/R \propto M^{1 - \beta}$, so it increases with mass provided $\beta < 1$. For a massive star, gravitational focusing dominates and mass accretion is a positive feedback process. Under these assumptions, the cross section for a massive star becomes
\begin{equation}
\label{RunawaySigma}
\sigma \approx \pi R^2 \frac{2 G M}{R V^2_\infty} \approx \frac{2 \pi G M^{1+\beta}}{V^2_\infty} 
\end{equation}
where the mass (and radius) of smaller stars has been neglected. Using Eq.~\ref{MassEvolution}, we obtain:
\begin{equation}
\label{RunawayAccretion}
\frac{dM}{dt} = m n \frac{2 \pi G M^{1 + \beta}}{V_\infty}
\end{equation}

At time $t = 0$ the (soon to be a) VMS had a mass $M_0$ and a radius $R_0$, so:
\begin{equation}
\label{RAbetter}
\frac{dM}{dt} = \frac{M_0}{\tau} {\left( \frac{M}{M_0} \right) }^{1+\beta}
\end{equation}
where
\begin{equation}
\label{tau}
\tau = \frac{\langle v \rangle}{2 \pi G n m R_0}
\end{equation}
so
\begin{equation}
\label{MassOfTime}
M(t) = M_0 { \left( 1 - \frac{\beta t}{\tau} \right)}^{-{1/\beta}}
\end{equation}
which at time
\begin{equation}
\label{divtime}
t_{DIV} = \frac{\tau}{\beta}
\end{equation}
formally diverges. A realistic value of $\beta$ is $1/2$, so the divergence takes place in a finite time. For $\beta \to 0$ the growth would instead become exponential.

A VMS can thus be formed if
\begin{equation}
\label{VMSisformed}
t_{DIV} < t_{MS} \approx 3 Myr
\end{equation}
where $t_{MS}$ is the typical lifetime of massive main-sequence stars, i.e. the time it takes for supernova explosions to take place.

Another way to express the collisional merging timescale $\tau$ is
\begin{equation}
\label{crossingtime}
\tau = \frac{R_C}{R_0} \tau_{cr}
\end{equation}
where $R_C$ is the radius of the young star cluster, $\tau_{cr}$ is its crossing time and $R_0$ is the typical radius of a star in the cluster. This naive timescale is actually too long for runaway merging to occur in a cluster with the typical crossing time and core radius observed today. To study the infall process and the subsequent runaway merging, thus going beyond the simple model described above, numerical approaches are needed, and they must necessarily take into account core-collapse and the very high densities it attains.

\subsection{IMBHs formed from merging of lower-mass black holes}
\cite{Mi2002} propose that dynamical interactions between binaries containing black-holes lead to repeated merging, which may result in the formation of a $\approx 10^3$ $M_{\odot}$ IMBH if the process continues for billions of years, as may be the case in old star clusters, such as galactic Globular Clusters (GCs). Three-body interactions are crucial for tightening the binaries, bringing about coalescence, but they risk to push the system out of the cluster, thus stopping the merging process. Therefore the seed black hole must be massive enough (at least $\approx 50$ $M_{\odot}$ according to \cite{Mi2002}) to avoid expulsion.

In a GC, the scale binding energy per unit mass is
\begin{equation}
\label{binding}
U_b = \alpha^2 \frac{G M_{GC}}{R_{GC}}
\end{equation}
where $\alpha$ is a dimensionless shape factor.
By setting
\begin{equation}
\label{vesc}
v_\infty = \sqrt{2 U_b}
\end{equation}
and, taking the typical values of a quite concentrated GC, $M_{GC} = {10}^{6}$ $M_{\odot}$, $R_{GC} = 1$ $pc$, we obtain
\begin{equation}
\label{vescnum}
v_\infty = \alpha \sqrt{\frac{2 G M_{GC}}{R_{GC}}} = \alpha \times 92 km/s
\end{equation}
whereas for a looser cluster, $M_{GC} = {10}^{5}$ $M_{\odot}$, $R_{GC} = 4$ $pc$, and
\begin{equation}
\label{vescnum2}
v_\infty = \alpha \times 21 km/s
\end{equation}
Thus, the typical escape velocity of a GC is about $\approx 50$ $km/s$.

Let us consider a BH binary, where the primary has mass $M$ and the secondary has mass $m \ll M$. In an interaction with a third object of mass $\approx m$ the binary hardens (i.e. increases its binding energy) by
\begin{equation}
\label{Hardening}
\Delta U = \eta \frac{m}{M} U
\end{equation}
where $\eta$ is typically $\approx 0.2$ \citep[][]{Qu1996}.
A fraction $\approx m/M$ of the energy $\Delta U$ lost by the orbital motion is transferred to the center-of-mass motion of the binary because of momentum conservation, so its overall kinetic energy increases by
\begin{equation}
\label{HardeningEnergy}
\Delta T = \eta \frac{m^2}{M^2} U
\end{equation}
To unbind the binary from the cluster, an increase of order the escape energy $E_{esc}$ is needed
\begin{equation}
\label{EscapeEnergy}
\Delta T = E_{esc} = \frac{1}{2} M v^2_{esc}
\end{equation}
so
\begin{equation}
\label{MinimumEnergyEq}
\eta \frac{m^2}{M^2} U = \frac{1}{2} M v^2_{esc}
\end{equation}
and solving for $U$ we obtain the minimum (in modulus) binding energy for which the binary is liable to be ejected due to a three-body interaction
\begin{equation}
\label{MinimumEnergy}
U_{ej} = \frac{M^3 v^2_{esc}}{2 m^2 \eta}
\end{equation}

$U_{ej}$ must now be compared to the minimum energy $U_{co}$ for which the binary coalesces, due to gravitational wave emission. The time it takes the binary to coalesce with this mechanism is \citep[][]{Mi2002}
\begin{equation}
\label{GWcoal}
t_c = \frac{5 c^5}{128 G^3} \frac{a^4}{\mu M^2} {\left( 1 - e^2\right)}^{7/2}
\end{equation}
for a binary orbit with semi-axis $a$, eccentricity $e$ and reduced mass $\mu = mM/(m + M) \approx m$.
Equation \ref{GWcoal} depends on orbital eccentricity, which varies in time, but we can take an average value $\langle e \rangle$. So, using Eq.~\ref{GWcoal}, we obtain
\begin{equation}
\label{Ucoquasi}
U = \frac{G M m}{a} = {\left[ \frac{5 G c^5}{128} {(1 - {\langle e \rangle}^2)}^{7/2} \right]}^{1/4} \frac{M^{1/2} m^{3/4}}{t^{1/4}_c}
\end{equation}
and $t_c$ must be shorter than the time $t_e$ between two subsequent three-body encounters, otherwise the binary gets a new three-body kick before being able to merge. For $t_c = t_e$ we obtain:
\begin{equation}
\label{Uco}
U_{co} = {\left[ \frac{5 G c^5}{128} {(1 - {\langle e \rangle}^2)}^{7/2} \right]}^{1/4} \frac{M^{1/2} m^{3/4}}{t^{1/4}_e}
\end{equation}

Now, if $U_{co} < U_{ej}$, merging can take place before ejection. Thus
\begin{equation}
\label{Condition}
 {\left[ \frac{5 G c^5}{128} {(1 - {\langle e \rangle}^2)}^{7/2} \right]}^{1/4} \frac{M^{1/2} m^{3/4}}{t^{1/4}_e} < \frac{M^3 v^2_{esc}}{2 m^2 \eta}
\end{equation}
and by setting
\begin{equation}
\label{k}
k = 2 \eta {\left[ \frac{5}{128} {\left(1 - {\langle e\rangle}^2\right)}^{7/2} \right]}^{1/4}
\end{equation}
we obtain
\begin{equation}
\label{ConditionOnM}
M > k^{2/5} {\left[ \frac{G c^5 m^{11}}{t_e v^8_{esc}} \right]}^{1/10}
\end{equation}
An alternative way to write this condition is
\begin{equation}
\label{ConditionOnMschw}
M > k^{2/5} {\left[ \frac{r_S/\lambda_e}{{(v_{esc}/c)}^7} \right]}^{1/10} m
\end{equation}
where $r_S$ is the Schwartzschild radius of the star of mass $m$, and the product $v_{esc} t_e$ was indicated with $\lambda_e$.
\cite{Mi2002} adopt $\eta \approx 0.2$ and $\langle e \rangle \approx 0.7$, obtaining $k \approx 0.1$. If $m = 10 M_{\odot}$, $v_{esc} = 50$ $km/s$ and $t_e = {10}^{6}$ $yr$, Eq.~\ref{ConditionOnM} becomes
\begin{equation}
\label{ConditionOnMNumbers}
M > 70 M_{\odot}
\end{equation}
This is approximately the threshold mass for IMBH growth through stellar-mass black hole merging. It is too large to come directly from stellar-mass black holes, so black-hole coalescence appears to need a seed black hole produced from some other process in order to work. On the other hand, once the accretion process is started, the IMBH grows quite fast, acquiring $\approx 10$ $M_{\odot}$ through merging events every $\approx {10}^6$ $yr$, which over the lifetime of a GC ($\approx {10}^{10}$ $yr$) can make it grow to $\approx {10}^5$ $M_\odot$ in mass, if enough stellar-mass black holes are present.

\section{Observational evidence of IMBH existence}
\label{Sect:CurrentIMBHevidence}
Evidence of the existence of IMBHs is still controversial. We will review two groups of phenomena: effects on the dynamics of stars in a cluster, and ultra-luminous X-ray emission. Several more phenomena are linked to IMBHs and could in principle yield a detection in the future: we sketch a discussion of these, but we do not add too much detail, as it would be premature.

\subsection{Stellar dynamics in clusters}
The best way to detect an IMBH in a star cluster would be to observe the orbits of stars bound to it.
This would allow to measure the black hole mass, as in the case of the super-massive black hole in the center of the Milky Way \citep[see][]{Gil2009, Gh2008, Gi2009b, Gh2005, Gh1998}, resulting in an uncertainty of $\approx 10\%$, which is remarkably small by astrophysical standards. Sadly, no study of this kind has been completed with a positive result in any star cluster. This is understandable if we recall that the surroundings of the black hole in the galactic center were observed for more than $10$ years, using state-of-the-art imaging technology ($10$ $m$ telescopes with adaptive optics at KECK). An observational program of this kind on Galactic GCs, which are approximately $200$, would literally require millennia. Moreover, cluster cores are extremely crowded (i.e. the images of several different stars will overlap) and the stars we need to observe are probably faint, main sequence stars, unless a red giant or an horizontal-branch star gets trapped in the IMBH potential well by chance.

The methods used to detect IMBHs must then be indirect, at least at this stage. Detection methods based on stellar dynamics can be divided between those which probe the zone of the cluster directly affected by the presence of the IMBH, known as the IMBH influence sphere, and those who do not need to. The former are more affected by crowding in the cluster core and by small number statistics, because the influence sphere usually contains few stars.

\subsubsection{The influence sphere}
Let the IMBH mass be $M_{BH}$ and the cluster mass be $M_{GC}$. The size of the region where the IMBH strongly affects the dynamics of stars may be calculated by equating the escape velocity $v_e$ at a distance $R_{INF}$ from the IMBH to the velocity dispersion $\sigma$ in the cluster core:
\begin{equation}
\label{InfSphere}
v_e = \sqrt{\frac{2 GM_{BH}}{R_{INF}}} = \sigma
\end{equation}
and
\begin{equation}
\label{InfSphereMore}
\sigma \approx \sqrt{\frac{G M_{GC}}{R_{GC}}}
\end{equation}
where $R_{GC}$ is a scale radius for the cluster. So, neglecting factors
\begin{equation}
\label{RSphere}
\frac{R_{INF}}{R_{GC}} \approx \frac{M_{BH}}{M_{GC}}
\end{equation}
for a typical globular star cluster, $R_{GC} \approx 1$ $pc$. If we take $M_{BH}/M_{GC} = {10}^{-2}$, then $R_{INF} = {10}^{-2}$ $pc$. The distance of the nearest globular star clusters from the Sun is on the order of several $kpc$. At a distance of $\approx 1$ $kpc$ the angular size of the influence sphere is $\approx 2$ $arcsec$. For nearby clusters it is thus feasible to probe the influence sphere with present-day telescopes (e.g. Hubble Space Telescope).
The volume of the influence sphere is
\begin{equation}
\label{VSphere}
R^3_{INF} \approx {\left( \frac{M_{BH}}{M_{GC}} \right)}^3 R^3_{GC}
\end{equation}
The number density of stars in the cluster core is a factor of $\eta \approx 10 \div 100$ larger than the average density, so the fraction of stars within the influence sphere is
\begin{equation}
\label{NSphere}
\frac{N_{INF}}{N_{GC}} = \eta {\left( \frac{M_{BH}}{M_{GC}} \right)}^3
\end{equation}
and given that $N_{GC} \approx {10}^5$ - ${10}^6$ for a typical globular star cluster we obtain
\begin{equation}
\label{NSphereNumber}
N_{INF} = 10 - {10}^2
\end{equation}
which is a small number if we hope to obtain a smooth density profile. This is one of the major hindrances encountered in revealing IMBHs using the dynamics of stars in the influence sphere. Still, some interesting results can be obtained, as shown in the following section.

\subsubsection{Evidence within the influence sphere}
\cite{Pe1972} and \cite{Ba1976} were amongst the first to predict the formation of density cusps in star clusters containing a central massive object.
Both find a phase space distribution function of stars near the central mass in the form
\begin{equation}
\label{BWCusp}
f(r, v) = K {(-E)}^p
\end{equation}
where $E$ is the energy per unit mass (assuming that all stars have the same stars)
\begin{equation}
\label{BWCuspEnergy}
E = \frac{v^2}{2} - \frac{G M_{BH}}{r} 
\end{equation}
and $r$ is the distance from the central object. The associated density profile is
\begin{equation}
\label{BWCuspIntegral}
n(r) = \int f d^3 v = 4 \pi K 2^{3/2} I_p {\left( \frac{GM_{BH}}{r} \right)}^{p + 3/2}
\end{equation}
where
\begin{equation}
\label{Ip}
I_p = \int^\infty_0 {\left( 1 - x^2 \right)}^p x^2 dx
\end{equation}
so a power-law cusp in the three-dimensional density profile is predicted, with exponent $p + 3/2$.
\cite{Ba1976} finds $p = 1/4$ based on an analytic argument. This result is confirmed by simulations based on different numerical approaches \citep[][]{Sh1985, Fr2002, Ba2004, Pr2004, Am2004}.
Observing a cusp in the core of a GC may thus be an indication of the presence of an IMBH, but some caveats apply. Cusps may form due to other physical processes, such as core collapse in a cluster containing stars with different masses \citep[][]{Tr2009}. On the other hand, a cusp may easily go undetected if the center of the cluster is incorrectly determined or if the resolution of the observations is limited. Notwithstanding these limitations \cite{No2006} and \cite{No2007} were able to reveal central luminosity density cusps in a fraction of the sample of GCs they observed using Hubble Space Telescope.

A better grip on the central dynamics of clusters can be obtained if spectroscopic (i.e. kinematic) data are also used.
\cite{No2008a} and \cite{No2008b} obtained kinematic information for the inner $5$ arcsec of $\omega$ Centauri using the Gemini telescopes, resulting in a claim to the detection of a $4 \times {10}^4$ $M_{\odot}$ black hole in its center. 
The point here is that the observed stellar velocities increase near the center (i.e. a velocity dispersion cusp is observed). A central rise in the velocity dispersion argues for the presence of a dark mass which deepens the cluster potential well in the center.
However, the kinematic information form spectroscopy is limited to line-of-sight velocities, which might not tell the whole story. A strong anisotropy in the central pressure tensor could account for the observed rise, as line-of-sight velocities could be much higher than velocities on the plane of the sky. Moreover, centering is still a delicate issue here, and the mechanisms which could generate a density cusp (see above) are likey to generate a velocity dispersion cusp as well.

At variance with the claim of \cite{No2008a}, \cite{Va2010} find an upper limit of $1.2 \times {10}^4$ $M_\odot$ to the mass of an IMBH in $\omega$ Centauri. A definitive solution to the puzzle has not been found yet.

\subsubsection{Evidence outside the influence sphere}
Evidence depending on resolving the influence sphere is limited by the low number of stars contained in it. Methods which rely on the global cluster structure to test for IMBH presence are free from to these shortcomings. The density profile of a GC, either three-dimensional or sky projected, is characterized by two characteristic radii and the overall luminosity.
The central part of the cluster is almost constant in density, up to a \emph{core radius}, $R_c$, from which the density starts to decline. So $R_c$ measures the size of the core of the cluster. The overall size of the cluster, instead, is characterized by $R_h$, the radius containing half of the cluster mass.

\cite{Tr2007} use N-body numerical simulations to find that, regardless of the initial conditions, clusters with an IMBH of about $1\%$ the total cluster mass and a realistic content of binary stars (about $10\%$) settle to a ratio of the (three-dimensional) core radius to the (three-dimensional) half-light radius of about
\begin{equation}
\label{RcRh}
\frac{R_c}{R_h} \approx 0.3
\end{equation}
after becoming sufficiently dynamically old. This is $\approx 3$ times as large as the ratio obtained for clusters without an IMBH. The core of a cluster containing an IMBH is larger in proportion to the overall size of the cluster.

Let us provide a justification of the above result from the physical point of view. We mentioned in the previous sections that a GC loses mass and energy as stars evaporate from it. Evaporation is due to two-body relaxation, i.e. stars exchanging momentum and energy in gravitational encounters. Two body relaxation promotes thermalization, i.e. the cluster phase space distribution function is driven towards a maxwellian over timescales of about one to tens of Gyr, depending on the cluster radius and number of stars.
A maxwellian, however, has a tail with arbitrarily high velocities. The stars ending up in such a tail must be ejected from the cluster, removing energy and mass from it. The evolution in time of the half-mass radius is influenced by this evaporation process as follows (the following analytical argument is loosely inspired by \cite{Ve1994} and \cite{He1965}).

The potential energy of the cluster is
\begin{equation}
\label{PotEnGeneric}
U = - \frac{1}{2} \sum_{i \neq j} \frac{G m_i m_j}{r_{ij}}
\end{equation}
where $m_i$ is the mass of the $i$-th star, and $r_{ij}$ is the distance between the $i$-th and the $j$-th star.
The virial theorem holds:
\begin{equation}
\label{VirialRespect}
2 T + U = 0
\end{equation}
so
\begin{equation}
\label{VirialEnergy}
E = T + U = \frac{U}{2}
\end{equation}
We introduce an effective radius of the cluster such that
\begin{equation}
\label{PotEnEffective}
U = - \frac{G M^2}{R_{eff}}
\end{equation}
where $M = \sum_i m_i$. Note that in general $R_{eff} \neq R_c \neq R_h$.
So
\begin{equation}
\label{REffective}
R_{eff} = \frac{2 M^2}{\sum_{i \neq j} \frac{m_i m_j}{r_{ij}}} = \frac{2 N^2}{\sum_{i \neq j} \frac{1}{r_{ij}}}
\end{equation}
where $N$ is the total number of stars. The last equality holds approximately for equal-mass stars.
The radius $R_{eff}$ can be very different from $R_h$, as if any $r_{ij} \to 0$, the denominator diverges and $R_{eff} \to 0$. In other words, if stars generate subclusters (e.g. binary or triple systems), their binding energy becomes extremely high, giving a significant contribution to the total binding energy of the cluster.
The ratio
\begin{equation}
\label{ReffoverRh}
\eta = \frac{R_{eff}}{R_h}
\end{equation}
measures the amount of subclustering present in the system: the larger $\eta$, the less subclustering is present.
From Eq.~\ref{PotEnEffective} and Eq.~\ref{VirialEnergy} we obtain
\begin{equation}
\label{TheMostImportantEquation}
\frac{\dot{E}}{E} =  2 \frac{\dot{M}}{M} - \frac{\dot{R}_{eff}}{R_{eff}}
\end{equation}
where the dot indicates a derivative with respect to time. Eq.~\ref{ReffoverRh} implies
\begin{equation}
\label{ReffoverRhDerivative}
\frac{\dot{R}_{eff}}{R_{eff}} = \frac{\dot{\eta}}{\eta} + \frac{\dot{R}_h}{R_h}
\end{equation}
so
\begin{equation}
\label{TheMostImportantEquationReloaded}
\frac{\dot{R_h}}{R_h} =  2 \frac{\dot{M}}{M} - \frac{\dot{E}}{E} - \frac{\dot{\eta}}{\eta}
\end{equation} 
The cluster loses mass due to evaporation, so $\dot{M}/M < 0$. It loses energy as well, but its binding energy is negative, so $\dot{E}/E > 0$. Then $2 \dot{M}/M - \dot{E}/E < 0$. If no substructure forms, $\dot{\eta} = 0$ and the only way the cluster has to satisfy the virial theorem and conserve energy is to globally shrink, reducing $R_h$. Shrinking increases the binding energy, compensating for the energy lost to evaporation. If instead $\dot{\eta}/\eta$ is negative, i.e. substructure forms, the cluster can keep energy balance without shrinking. Lower values of $\eta$ correspond to the formation of hierachical subclusters, so binary stars form (or harden) in order to increase the cluster binding energy while allowing for a constant or even expanding $R_h$.
This makes physical sense, as binaries can form only via three-body interactions (the third star acts as an energy sink). Three body interactions need an high density of stars to happen in significant numbers. An initial shrinkage of the cluster makes the central density rise, preparing the conditions for three-body interactions to take place and for binaries to be generated, or hardened if they are already present.

In simulations, the IMBH is usually found in a binary with a stellar-mass black-hole, so it acts as a source of energy, giving a significant contribution to the substructure term $\dot{\eta}/\eta$. Sources of energy in the form of substructure (several binaries or an high-mass binary with the IMBH as principal) can halt the collapse of the cluster and sustain a slight expansion in $R_h$ as well as a large core with high $R_c/R_h$. This explains the result expressed by Eq.~\ref{RcRh}.

Observationally, $R_c/R_h$ can be used as a very preliminary method to select GCs which are eligible as IMBH hosts, but several difficulties are present. Sky-projected quantities only are measurable in an observed GC. The half-mass radius is not directly observable, as only luminous matter can be used to compute it, and the dark remnants (neutron stars, stellar mass black holes, white dwarfs) are typically more massive than visible stars and tend to segregate to the center of GCs, thus leading to an overestimate of $R_h$ as in general the light profile does not trace the mass profile of the cluster. This is a major hindrance, as the relative amount of stellar-mass black holes and neutron stars with respect to the visible stellar component is poorly constrained.

\subsection{Ultra Luminous X-ray sources (ULXs)}
\label{ULXs}
\subsubsection{Eddington limit}
A source of radiation of mass $M$ held together by its own gravity cannot have an arbitrarily high luminosity. A particle of mass $m$ and charge $q$ at distance $r$ from the source feels gravitational attraction to the source
\begin{equation}
\label{attraction}
F_g = \frac{G M m}{r^2}
\end{equation}
and repulsion from it due to radiation pressure
\begin{equation}
\label{radpressure}
F_r = \frac{\sigma L}{4 \pi c r^2}
\end{equation}
where $\sigma = \sigma(q)$ is the cross section of the particle to the radiation from the source, and $L$ is the source luminosity.
Equation \ref{radpressure} holds only in the case of spherically symmetric emission.

By requiring that $F_r < F_g$ so the particle is globally attracted to the source we obtain
\begin{equation}
\label{eddington}
L < \frac{G M m 4 \pi c}{\sigma} = L_{EDD}
\end{equation}
so for any mass $M$ of the source there exists a critical luminosity $L_{EDD}$ at which radiation pressure would destroy the source.
We can take $\sigma_{Te}$ (the Thompson cross section) as the electron cross section
\begin{equation}
\label{ThompsonElectron}
\sigma_{Te} = \frac{8 \pi}{3} r^2_e = \frac{8 \pi q^4_e}{3 m^2_e c^4}
\end{equation}
where $r_e$ is the \emph{classical electron radius}, $q_e$ and $m_e$ the electron's charge and mass.
Instead for a proton we obtain
\begin{equation}
\label{ThompsonProton}
\sigma_{Tp} = \frac{8 \pi q^4_e}{3 m^2_p c^4}.
\end{equation}
Thus $\sigma_{Tp} \ll \sigma_{Te}$, because $m_e \ll m_p$.
As an estimate for hydrogen, we use $m_p = 1.67 \times {10}^{-24}$ $g$ for $m$ and $\sigma_{Te} = 6.65 \times {10}^{-25}$ ${cm}^2$ for $\sigma$, obtaining
\begin{equation}
\label{eddingtonnumbers}
L_{EDD} = \frac{M}{M_{\odot}} \times 1.26 \times {10}^{38} erg/s
\end{equation}
as the maximum allowed luminosity for the source not to be destroyed by its own radiation pressure.

\subsubsection{Observing and explaining ULXs}
An unresolved X-ray source (i.e. a source that appears point-like in observations) is defined as ultra-luminous if
\begin{equation}
\label{ULXdef}
L_X > {10}^{39} erg/s
\end{equation}
i.e. if its luminosity in the X-ray band exceeds the Eddington luminosity of an $\approx 10 M_{\odot}$ mass.
The centers of galaxies often display X-ray activity, but they are usually excluded from this class of objects. Thus only those sources which are off-center with respect to the parent galaxy are counted among the ULXs.

High-luminosity X-ray sources began to be discovered in galaxies thanks to the \emph{Einstein Observatory} mission in 1978 \citep[][]{Fab1989}. ROSAT, ASCA, \emph{Chandra} and \emph{XMM-Newton} followed, resulting in a wealth of data on the X-ray sky. ULXs are widespread in nearby galaxies, but a satisfactory understanding of the underlying physical mechanisms has not been attained yet \citep[][]{Fab2006}. To achieve luminosities as high as ${10}^{39}$ $erg/s$, the only plausible physical mechanism proposed so far is accretion of matter onto a compact object. However, by definition, ULXs are impossible to explain by isotropic, sub-Eddington emission from an accreting stellar mass black hole of $M \lesssim 10 M_{\odot}$.

The only remaining viable explanations are that either ULXs are beamed sources or somehow achieve super-Eddington emission, or a new class of accreting objects, that is IMBHs, is the solution to this observational riddle. We point the reader to \cite{Col2005} for a review of the observational evidence supporting the latter point of view. In the following we briefly review the most important piece of evidence, that is spectroscopic data.

\subsubsection{Spectroscopy and the MCD model}
The Multi Color Disk (MCD) model was first used to fit the X-ray spectra of Galactic sources emitting at lower luminosities than ${10}^{39}$ $erg/s$. It stems essentially from the superposition of a blackbody spectrum and the so-called multi color disk spectrum \citep[][]{Mi1984, Ta1994}. Galactic X-ray sources are neutron stars or stellar-mass black holes accreting mass from a companion star. The accreting matter forms a disk orbiting the neutron star or black hole and increases its temperature as it nears its surface. The blackbody component corresponds to the surface of the accreting object, and the multi color disk component is a sum of blackbody sources at different temperatures intended to model the accretion disk.

A particle of mass $m$ in a newtonian orbit of semi-axis $r$ around the accretor mass $M$ has an energy
\begin{equation}
\label{keplerianenergy}
E = -\frac{GMm}{2 r}
\end{equation}
so, a change of its semi-axis by an amount $dr$ influences the energy as
\begin{equation}
\label{kepleriandenergy}
dE = \frac{GMm}{2r^2} dr
\end{equation}
Such a change can be brought about by a perturbing force, most likely friction due to gas viscosity in our case.

The energy $E_d$ deposited per unit time by the accreting matter in the region between $r$ and $r + dr$ scales as $E_d \propto dr/r^2$. The energy $E_T$ radiated per unit time, on the other hand, scales with the surface of the emitting region, i.e. is proportional to $r dr T^4$, so
\begin{equation}
\label{thermalequilibrium}
\frac{dr}{r^2} \propto r dr T^4
\end{equation}
which, integrating, yields
\begin{equation}
\label{Tscaling}
T \propto r^{-3/4}
\end{equation}
so for a face-on disk at distance $D$ from the observer we obtain a spectrum
\begin{equation}
\label{MCDintegral}
\frac{1}{D^2} \int_{r_I}^{r_O} I(E, T) 2 \pi r dr = \frac{8 \pi r^2_I}{3 D^2 T_I} \int_{T_O}^{T_I} {\left( \frac{T}{T_I} \right)}^{-11/3} I(E, T) dT
\end{equation}
where
\begin{equation}
\label{PlanckDistribution}
I(E, T) dE =  A \frac{E^3 dE}{(e^{E/kT} - 1)}
\end{equation}
is the Planck distribution, $r_I$ is the innermost radius of the disk, nearest to the accreting object, $r_O$ the outermost radius and $T_I$ and $T_O$ the corresponding temperatures at these radii. Notice that $r_O \gg r_I$ and $T_I > T_O$.

The parameter $T_I$, which is thus obtained by fitting the spectrum, depends on the mass of the accreting object \citep[][]{Ma2000}:
\begin{equation}
\label{MassScalingTI}
T_I \propto M^{-1/4}
\end{equation}
so lower temperatures of the inner edge, correspond to higher masses of the accreting object.

So high-luminosity ULXs with spectra that are well fit by a MCD model with low $T_I$ are good IMBH candidates.
Typical temperatures obtained from Galactic X-ray sources associated with stellar-mass accreting objects are in the $0.4$ - $1.0$ $keV$ range, so the expected temperature for an object $\approx 100$ times more massive will range from $0.1$  to $0.3$ $keV$.
\cite{Col2005} (Table $1$) report several ULXs having $T_I$ in this range and a luminosity in excess of ${10}^{40}$ $erg/s$, which are potential IMBH candidates.

\subsubsection{ULXs and star-forming regions}
As a last remark on ULXs, we notice that they appear to be associated to active star-forming regions \citep[][]{Hu2003, Sw2004, Li2005}, in particular interacting galaxy pairs: e.g. the Antennae (NGC $4038$/$4039$) \citep[][]{Fa2001} and the interacting galaxy pair NGC $4485$/$4490$ \citep[][]{Ro2002}. If ULXs are really associated to IMBHs this feature would support the runaway merging model of IMBH origin, as young clusters with high-mass main sequence stars are found virtually only in star-forming regions.

\subsection{Other phenomena possibly related to IMBHs}
Other phenomena have been regarded as possible signs of the presence of an IMBH in a star cluster. Some of them are reviewed in the following, with the caveat that the listing may be incomplete due to the large amount of work published on the subject.

\subsubsection{Millisecond pulsar timing}
\cite{Da2002} use the timing of three Milli-Second Pulsars (MSPs) in the GC NGC $6752$ to obtain lower limits to their line-of-sight acceleration due to the cluster potential. This implies a lower limit of about $10$ $M_\odot/L_\odot$ for the central mass-to-light ratio, equivalent to $1.3 \times {10}^4$ $M_\odot$ in dark objects within the innermost $0.15$ $pc$ around the cluster center. IMBH of this mass would be a solution, but such a central increase in the mass-to-light ratio can be explained away by dark remnants (stellar-mass black holes and neutron stars) which undergo mass-segregation and accumulate in the center of the cluster potential well.

\subsubsection{High-velocity stars}
\cite{Tr2007} shows, among other results, that a population of hard binaries interacting with an IMBH can produce several stars which escape the cluster with a velocity of hundreds of $km/s$, much higher than typical cluster escape velocities.
\cite{Hi1988} proposed to use high-velocity stars (possibly) ejected from the galactic center as a test for the presence of a super-massive black hole. \cite{Yu2003} estimates the number and the velocity of high-velocity stars produced by various mechanisms (gravitational encounters of single stars, tidal breakup of a binary star, ejection by a binary black hole) analytically in quantitative detail. In principle these estimates can be adapted to star clusters, yielding a potential method for detecting IMBHs.

\appendix
\section*{Acknowledgments}
I wish to thank Jarah Evslin for organizing the school and asking me to participate with these lectures.
Without the help of Jasmina Selmic, the croatian version of the abstract would have been based on google translate.
While that would have been a perfect way to make any croatian reader laugh, I wish to thank her for her
translation efforts. Any spelling mistake is still my own responsibility, though, as I typed the text without much knowledge of the language. Last but not least, I thank Raffaele Savelli and his car for taking us from Trieste to Trpanj.

\bibliographystyle{aa}
\bibliography{lecturenotes}

\begin{thebibliography}{72}
\expandafter\ifx\csname natexlab\endcsname\relax\def\natexlab#1{#1}\fi

\bibitem[{{Abbott}(1980)}]{Ab1980}
{Abbott}, D.~C. 1980, \apj, 242, 1183

\bibitem[{{Amaro-Seoane} {et~al.}(2004){Amaro-Seoane}, {Freitag}, \&
  {Spurzem}}]{Am2004}
{Amaro-Seoane}, P., {Freitag}, M., \& {Spurzem}, R. 2004, \mnras, 352, 655

\bibitem[{{Bahcall} \& {Wolf}(1976)}]{Ba1976}
{Bahcall}, J.~N. \& {Wolf}, R.~A. 1976, \apj, 209, 214

\bibitem[{{Baumgardt} {et~al.}(2004){Baumgardt}, {Makino}, \&
  {Ebisuzaki}}]{Ba2004}
{Baumgardt}, H., {Makino}, J., \& {Ebisuzaki}, T. 2004, \apj, 613, 1133

\bibitem[{{Bromm} {et~al.}(2002){Bromm}, {Coppi}, \& {Larson}}]{Br2002}
{Bromm}, V., {Coppi}, P.~S., \& {Larson}, R.~B. 2002, \apj, 564, 23

\bibitem[{{Bromm} \& {Larson}(2004)}]{Br2004}
{Bromm}, V. \& {Larson}, R.~B. 2004, \araa, 42, 79

\bibitem[{{Bulik}(2007)}]{Bu2007}
{Bulik}, T. 2007, \nat, 449, 799

\bibitem[{{Cassinelli}(1979)}]{Ca1979}
{Cassinelli}, J.~P. 1979, \araa, 17, 275

\bibitem[{{Castor} {et~al.}(1975){Castor}, {Abbott}, \& {Klein}}]{Ca1975}
{Castor}, J.~I., {Abbott}, D.~C., \& {Klein}, R.~I. 1975, \apj, 195, 157

\bibitem[{{Colbert} \& {Miller}(2005)}]{Col2005}
{Colbert}, E.~J.~M. \& {Miller}, M.~C. 2005, in The Tenth Marcel Grossmann
  Meeting. On recent developments in theoretical and experimental general
  relativity, gravitation and relativistic field theories, ed. {M.~Novello,
  S.~Perez Bergliaffa, \& R.~Ruffini}, 530

\bibitem[{{Colgate}(1967)}]{Co1967}
{Colgate}, S.~A. 1967, \apj, 150, 163

\bibitem[{Creighton \& Price(2008)}]{Cr2008}
Creighton, T. \& Price, R.~H. 2008, Scholarpedia, 3, 4277

\bibitem[{{D'Amico} {et~al.}(2002){D'Amico}, {Possenti}, {Fici}, {Manchester},
  {Lyne}, {Camilo}, \& {Sarkissian}}]{Da2002}
{D'Amico}, N., {Possenti}, A., {Fici}, L., {et~al.} 2002, \apjl, 570, L89

\bibitem[{{Ebisuzaki} {et~al.}(2001){Ebisuzaki}, {Makino}, {Tsuru}, {Funato},
  {Portegies Zwart}, {Hut}, {McMillan}, {Matsushita}, {Matsumoto}, \&
  {Kawabe}}]{Eb2001}
{Ebisuzaki}, T., {Makino}, J., {Tsuru}, T.~G., {et~al.} 2001, \apjl, 562, L19

\bibitem[{{Fabbiano}(1989)}]{Fab1989}
{Fabbiano}, G. 1989, \araa, 27, 87

\bibitem[{{Fabbiano}(2006)}]{Fab2006}
{Fabbiano}, G. 2006, \araa, 44, 323

\bibitem[{{Fabbiano} {et~al.}(2001){Fabbiano}, {Zezas}, \& {Murray}}]{Fa2001}
{Fabbiano}, G., {Zezas}, A., \& {Murray}, S.~S. 2001, \apj, 554, 1035

\bibitem[{{Ferrarese} \& {Ford}(2005)}]{Fe2005}
{Ferrarese}, L. \& {Ford}, H. 2005, Space Science Reviews, 116, 523

\bibitem[{{Figer} {et~al.}(1998){Figer}, {Najarro}, {Morris}, {McLean},
  {Geballe}, {Ghez}, \& {Langer}}]{Fi1998}
{Figer}, D.~F., {Najarro}, F., {Morris}, M., {et~al.} 1998, \apj, 506, 384

\bibitem[{{Fregeau} {et~al.}(2006){Fregeau}, {Larson}, {Miller},
  {O'Shaughnessy}, \& {Rasio}}]{Fr2006}
{Fregeau}, J.~M., {Larson}, S.~L., {Miller}, M.~C., {O'Shaughnessy}, R., \&
  {Rasio}, F.~A. 2006, \apjl, 646, L135

\bibitem[{{Freitag} \& {Benz}(2002)}]{Fr2002}
{Freitag}, M. \& {Benz}, W. 2002, \aap, 394, 345

\bibitem[{{Freitag} {et~al.}(2006){Freitag}, {Rasio}, \& {Baumgardt}}]{Fr2006a}
{Freitag}, M., {Rasio}, F.~A., \& {Baumgardt}, H. 2006, \mnras, 368, 121

\bibitem[{{Ghez} {et~al.}(2003){Ghez}, {Duch{\^e}ne}, {Matthews}, {Hornstein},
  {Tanner}, {Larkin}, {Morris}, {Becklin}, {Salim}, {Kremenek}, {Thompson},
  {Soifer}, {Neugebauer}, \& {McLean}}]{Gh2003}
{Ghez}, A.~M., {Duch{\^e}ne}, G., {Matthews}, K., {et~al.} 2003, \apjl, 586,
  L127

\bibitem[{{Ghez} {et~al.}(1998){Ghez}, {Klein}, {Morris}, \&
  {Becklin}}]{Gh1998}
{Ghez}, A.~M., {Klein}, B.~L., {Morris}, M., \& {Becklin}, E.~E. 1998, \apj,
  509, 678

\bibitem[{{Ghez} {et~al.}(2005){Ghez}, {Salim}, {Hornstein}, {Tanner}, {Lu},
  {Morris}, {Becklin}, \& {Duch{\^e}ne}}]{Gh2005}
{Ghez}, A.~M., {Salim}, S., {Hornstein}, S.~D., {et~al.} 2005, \apj, 620, 744

\bibitem[{{Ghez} {et~al.}(2008){Ghez}, {Salim}, {Weinberg}, {Lu}, {Do}, {Dunn},
  {Matthews}, {Morris}, {Yelda}, {Becklin}, {Kremenek}, {Milosavljevic}, \&
  {Naiman}}]{Gh2008}
{Ghez}, A.~M., {Salim}, S., {Weinberg}, N.~N., {et~al.} 2008, \apj, 689, 1044

\bibitem[{{Gill} {et~al.}(2008){Gill}, {Trenti}, {Miller}, {van der Marel},
  {Hamilton}, \& {Stiavelli}}]{Gi2008}
{Gill}, M., {Trenti}, M., {Miller}, M.~C., {et~al.} 2008, \apj, 686, 303

\bibitem[{{Gillessen} {et~al.}(2009{\natexlab{a}}){Gillessen}, {Eisenhauer},
  {Fritz}, {Bartko}, {Dodds-Eden}, {Pfuhl}, {Ott}, \& {Genzel}}]{Gi2009b}
{Gillessen}, S., {Eisenhauer}, F., {Fritz}, T.~K., {et~al.} 2009{\natexlab{a}},
  \apjl, 707, L114

\bibitem[{{Gillessen} {et~al.}(2009{\natexlab{b}}){Gillessen}, {Eisenhauer},
  {Trippe}, {Alexander}, {Genzel}, {Martins}, \& {Ott}}]{Gil2009}
{Gillessen}, S., {Eisenhauer}, F., {Trippe}, S., {et~al.} 2009{\natexlab{b}},
  \apj, 692, 1075

\bibitem[{{Glover}(2005)}]{Gl2005}
{Glover}, S. 2005, Space Science Reviews, 117, 445

\bibitem[{{Glover} {et~al.}(2008){Glover}, {Clark}, {Greif}, {Johnson},
  {Bromm}, {Klessen}, \& {Stacy}}]{Gl2008}
{Glover}, S.~C.~O., {Clark}, P.~C., {Greif}, T.~H., {et~al.} 2008, in IAU
  Symposium, Vol. 255, IAU Symposium, ed. {L.~K.~Hunt, S.~Madden, \&
  R.~Schneider}, 3--17

\bibitem[{{Heger} \& {Woosley}(2002)}]{He2002}
{Heger}, A. \& {Woosley}, S.~E. 2002, \apj, 567, 532

\bibitem[{{H{\'e}non}(1965)}]{He1965}
{H{\'e}non}, M. 1965, Annales d'Astrophysique, 28, 62

\bibitem[{{Hills}(1988)}]{Hi1988}
{Hills}, J.~G. 1988, \nat, 331, 687

\bibitem[{{Hoyle}(1953)}]{Ho1953}
{Hoyle}, F. 1953, \apj, 118, 513

\bibitem[{{Humphrey} {et~al.}(2003){Humphrey}, {Fabbiano}, {Elvis}, {Church},
  \& {Ba{\l}uci{\'n}ska-Church}}]{Hu2003}
{Humphrey}, P.~J., {Fabbiano}, G., {Elvis}, M., {Church}, M.~J., \&
  {Ba{\l}uci{\'n}ska-Church}, M. 2003, \mnras, 344, 134

\bibitem[{{Kawaguchi} {et~al.}(2008){Kawaguchi}, {Kawasaki}, {Takayama},
  {Yamaguchi}, \& {Yokoyama}}]{Ka2008}
{Kawaguchi}, T., {Kawasaki}, M., {Takayama}, T., {Yamaguchi}, M., \&
  {Yokoyama}, J. 2008, \mnras, 388, 1426

\bibitem[{{Kormendy} \& {Richstone}(1995)}]{Ko1995}
{Kormendy}, J. \& {Richstone}, D. 1995, \araa, 33, 581

\bibitem[{{Liu} \& {Bregman}(2005)}]{Li2005}
{Liu}, J. \& {Bregman}, J.~N. 2005, \apjs, 157, 59

\bibitem[{{Low} \& {Lynden-Bell}(1976)}]{Lo1976}
{Low}, C. \& {Lynden-Bell}, D. 1976, \mnras, 176, 367

\bibitem[{{Lucy} \& {Solomon}(1970)}]{Lu1970}
{Lucy}, L.~B. \& {Solomon}, P.~M. 1970, \apj, 159, 879

\bibitem[{{Madau} \& {Rees}(2001)}]{Ma2001}
{Madau}, P. \& {Rees}, M.~J. 2001, \apjl, 551, L27

\bibitem[{{Makishima} {et~al.}(2000){Makishima}, {Kubota}, {Mizuno}, {Ohnishi},
  {Tashiro}, {Aruga}, {Asai}, {Dotani}, {Mitsuda}, {Ueda}, {Uno}, {Yamaoka},
  {Ebisawa}, {Kohmura}, \& {Okada}}]{Ma2000}
{Makishima}, K., {Kubota}, A., {Mizuno}, T., {et~al.} 2000, \apj, 535, 632

\bibitem[{{McClintock} \& {Remillard}(2003)}]{Mc2003}
{McClintock}, J.~E. \& {Remillard}, R.~A. 2003, ArXiv Astrophysics e-prints

\bibitem[{{Mestel}(1965{\natexlab{a}})}]{Me1965}
{Mestel}, L. 1965{\natexlab{a}}, \qjras, 6, 161

\bibitem[{{Mestel}(1965{\natexlab{b}})}]{Me1965b}
{Mestel}, L. 1965{\natexlab{b}}, \qjras, 6, 265

\bibitem[{{Miller}(2002)}]{Mi2002b}
{Miller}, M.~C. 2002, \apj, 581, 438

\bibitem[{{Miller} \& {Hamilton}(2002)}]{Mi2002}
{Miller}, M.~C. \& {Hamilton}, D.~P. 2002, \mnras, 330, 232

\bibitem[{{Mitsuda} {et~al.}(1984){Mitsuda}, {Inoue}, {Koyama}, {Makishima},
  {Matsuoka}, {Ogawara}, {Suzuki}, {Tanaka}, {Shibazaki}, \& {Hirano}}]{Mi1984}
{Mitsuda}, K., {Inoue}, H., {Koyama}, K., {et~al.} 1984, \pasj, 36, 741

\bibitem[{{Norman}(2008)}]{No2008}
{Norman}, M.~L. 2008, in American Institute of Physics Conference Series, Vol.
  990, First Stars III, ed. {B.~W.~O'Shea \& A.~Heger}, 3--15

\bibitem[{{Noyola} \& {Gebhardt}(2006)}]{No2006}
{Noyola}, E. \& {Gebhardt}, K. 2006, \aj, 132, 447

\bibitem[{{Noyola} \& {Gebhardt}(2007)}]{No2007}
{Noyola}, E. \& {Gebhardt}, K. 2007, \aj, 134, 912

\bibitem[{{Noyola} {et~al.}(2008{\natexlab{a}}){Noyola}, {Gebhardt}, \&
  {Bergmann}}]{No2008a}
{Noyola}, E., {Gebhardt}, K., \& {Bergmann}, M. 2008{\natexlab{a}}, in IAU
  Symposium, Vol. 246, IAU Symposium, ed. {E.~Vesperini, M.~Giersz, \&
  A.~Sills}, 341--345

\bibitem[{{Noyola} {et~al.}(2008{\natexlab{b}}){Noyola}, {Gebhardt}, \&
  {Bergmann}}]{No2008b}
{Noyola}, E., {Gebhardt}, K., \& {Bergmann}, M. 2008{\natexlab{b}}, \apj, 676,
  1008

\bibitem[{{Pasquato} {et~al.}(2009){Pasquato}, {Trenti}, {De Marchi}, {Gill},
  {Hamilton}, {Miller}, {Stiavelli}, \& {van der Marel}}]{Pa2009}
{Pasquato}, M., {Trenti}, M., {De Marchi}, G., {et~al.} 2009, \apj, 699, 1511

\bibitem[{{Peebles}(1972)}]{Pe1972}
{Peebles}, P.~J.~E. 1972, \apj, 178, 371

\bibitem[{{Portegies Zwart} {et~al.}(2004){Portegies Zwart}, {Baumgardt},
  {Hut}, {Makino}, \& {McMillan}}]{Po2004}
{Portegies Zwart}, S.~F., {Baumgardt}, H., {Hut}, P., {Makino}, J., \&
  {McMillan}, S.~L.~W. 2004, \nat, 428, 724

\bibitem[{{Preto} {et~al.}(2004){Preto}, {Merritt}, \& {Spurzem}}]{Pr2004}
{Preto}, M., {Merritt}, D., \& {Spurzem}, R. 2004, \apjl, 613, L109

\bibitem[{{Quinlan}(1996)}]{Qu1996}
{Quinlan}, G.~D. 1996, New Astronomy, 1, 35

\bibitem[{{Roberts} {et~al.}(2002){Roberts}, {Warwick}, {Ward}, \&
  {Murray}}]{Ro2002}
{Roberts}, T.~P., {Warwick}, R.~S., {Ward}, M.~J., \& {Murray}, S.~S. 2002,
  \mnras, 337, 677

\bibitem[{{Schneider} {et~al.}(2002){Schneider}, {Ferrara}, {Natarajan}, \&
  {Omukai}}]{Sc2002}
{Schneider}, R., {Ferrara}, A., {Natarajan}, P., \& {Omukai}, K. 2002, \apj,
  571, 30

\bibitem[{{Shapiro}(1985)}]{Sh1985}
{Shapiro}, S.~L. 1985, in IAU Symposium, Vol. 113, Dynamics of Star Clusters,
  ed. {J.~Goodman \& P.~Hut}, 373--412

\bibitem[{{Swartz} {et~al.}(2004){Swartz}, {Ghosh}, {Tennant}, \&
  {Wu}}]{Sw2004}
{Swartz}, D.~A., {Ghosh}, K.~K., {Tennant}, A.~F., \& {Wu}, K. 2004, \apjs,
  154, 519

\bibitem[{{Takano} {et~al.}(1994){Takano}, {Mitsuda}, {Fukazawa}, \&
  {Nagase}}]{Ta1994}
{Takano}, M., {Mitsuda}, K., {Fukazawa}, Y., \& {Nagase}, F. 1994, \apjl, 436,
  L47

\bibitem[{{Tanaka} \& {Haiman}(2009)}]{Ta2009}
{Tanaka}, T. \& {Haiman}, Z. 2009, \apj, 696, 1798

\bibitem[{{Trenti} {et~al.}(2007){Trenti}, {Ardi}, {Mineshige}, \&
  {Hut}}]{Tr2007}
{Trenti}, M., {Ardi}, E., {Mineshige}, S., \& {Hut}, P. 2007, \mnras, 374, 857

\bibitem[{{Trenti} {et~al.}(2010){Trenti}, {Vesperini}, \& {Pasquato}}]{Tr2009}
{Trenti}, M., {Vesperini}, E., \& {Pasquato}, M. 2010, \apj, 708, 1598

\bibitem[{{van der Marel} \& {Anderson}(2010)}]{Va2010}
{van der Marel}, R.~P. \& {Anderson}, J. 2010, \apj, 710, 1063

\bibitem[{{Vesperini} \& {Chernoff}(1994)}]{Ve1994}
{Vesperini}, E. \& {Chernoff}, D.~F. 1994, \apj, 431, 231

\bibitem[{{Vink} {et~al.}(2001){Vink}, {de Koter}, \& {Lamers}}]{Vi2001}
{Vink}, J.~S., {de Koter}, A., \& {Lamers}, H.~J.~G.~L.~M. 2001, \aap, 369, 574

\bibitem[{{Volonteri} {et~al.}(2003){Volonteri}, {Haardt}, \& {Madau}}]{Vo2003}
{Volonteri}, M., {Haardt}, F., \& {Madau}, P. 2003, \apj, 582, 559

\bibitem[{{Yu} \& {Tremaine}(2003)}]{Yu2003}
{Yu}, Q. \& {Tremaine}, S. 2003, \apj, 599, 1129

\end{thebibliography}

\end{document}